\documentclass[prl,superscriptaddress,showpacs,twocolumn,epsf]{revtex4}
\usepackage{epsfig}

\begin{document}


\title{Quantum advantage does not survive in the presence of a corrupt source: Optimal strategies in simultaneous move games}
\author{\c{S}ahin Kaya \"Ozdemir}
\author{Junichi Shimamura}
\affiliation{SORST Research Team for Interacting Carrier Electronics,
CREST Research Team for Photonic Quantum Information,
The Graduate University for Advanced Studies (SOKENDAI), Hayama, Kanagawa 240-0193, Japan }
\author{Nobuyuki
Imoto} \affiliation{SORST Research Team for Interacting Carrier Electronics,
CREST Research Team for Photonic Quantum Information,
The Graduate University for Advanced Studies
(SOKENDAI), Hayama, Kanagawa 240-0193, Japan } \affiliation{NTT
Basic Research Laboratories, NTT Corporation, 3-1 Morinosato
Wakamiya, Atsugi, Kanagawa 243-0198, Japan}

\begin{abstract}
Effects of a corrupt source on the dynamics of simultaneous move
strategic games are analyzed both for classical and quantum
settings. The corruption rate dependent changes in the payoffs and
strategies of the players are observed. It is shown that there is
a critical corruption rate at which the players lose their quantum
advantage, and that the classical strategies are more robust to
the corruption in the source. Moreover, it is understood that the
information on the corruption rate of the source may help the
players choose their optimal strategy for resolving the dilemma
and increase their payoffs. The study is carried out in two
different corruption scenarios for Prisoner's Dilemma, Samaritan's
Dilemma, and Battle of Sexes.
\end{abstract}
\pacs{02.50.Le, 03.67.-a}
\date{\today}
\pagestyle{plain} \pagenumbering{arabic} \maketitle

{\bf\textit{a. Introduction:}}  Classical game theory has a very
general scope, encompassing questions and situations that are
basic to all of the social sciences \cite{Books}. There are three
main ingredients of a game which is to be a model for real life
situations \cite{Books}: The first of these is the rational
players (decision makers) who share a common knowledge. The second
is the strategy set which contains the feasible actions the
players can take, and the third one is the payoff which are given
to the players as their profit or benefit when they apply a
specific action from their strategy set. When rational players
interact in a game, they will not play dominated strategies, but
will search for an equilibrium. One of the important concepts in
game theory is that of Nash equilibrium (NE) in which each
player's choice of action is the best response to the actions
taken by the other players. In an NE, no player can increase his
payoff by unilaterally changing her action. While the existence of
a unique NE makes it easier for the players to choose their
action, the existence of multiple NE's avoids the sharp decision
making process because the players become indifferent between
them. In pure strategies, the type and the number of NE's in a
game depend on the game. However, due to von Neumann there is at
least one NE when the same game is played with mixed strategies
\cite{Books,Books1}. Classical game theory has been successfully
tested in decision making processes encountered in real-life
situations ranging from economics to international relations. By
studying and applying the principles of game theory, one can
formulate effective strategies, predict the outcome of strategic
situations, select or design the best game to be played, and
determine competitor behavior, as well as the optimal strategy.

\begin{figure}[h] \epsfxsize=7cm
\epsfbox{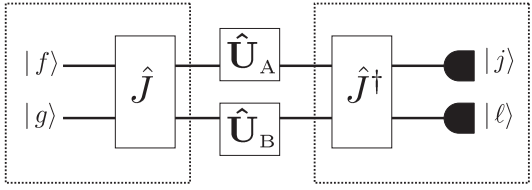}\caption[]{Physical scheme for playing
the quantum version of classical games. The operations inside the
dotted boxes are performed by the referee.}\label{Fig:scheme}
\end{figure}

In recent years, there have been great efforts to apply the
quantum mechanical toolbox in the design and analysis of games
\cite{Eisert1,Du,Eisert2,Meyer,Du2,Johnson3,Ben,Flitney2,Iqbal,Johnson1,Johnson2,Ozdemir}.
As it was the same in other fields such as communication and
computation, quantum mechanics introduced novel effects into game
theory, too.  It has proved to have the potential to affect our
way of thinking when approaching to games and game modelling.
Using the physical scheme proposed by Eisert {\it et al.} (see
Fig.\ref{Fig:scheme}) \cite{Eisert1}, it has been shown in several
games that the dilemma existing in the original game can be
resolved by using the paradigm of quantum mechanics
\cite{Eisert1,Du,Eisert2,Meyer,Du2,Johnson3,Ben,Flitney2,Iqbal,Johnson1,Johnson2,Ozdemir}.
It has also been shown that when one of the players chooses
quantum strategies while the other is restricted to classical
ones, the player with quantum strategies can always receive better
payoff if they share a maximally entangled state \cite{Text2a}.

Quantum systems are easily affected by their environment, and
physical schemes are usually far from ideal in practical
situations. Therefore, it is important to study whether the
advantage of the players arising from the quantum strategies and
the shared entanglement survive in the presence of noise or
non-ideal components in the physical scheme. In this paper, we
consider a corrupt source and analyze its effect on the payoffs
and strategies of the players. We search answers for the following
two questions: (i) Is there a critical corruption rate above which
the players cannot maintain their quantum advantage if they are
unaware of the action of the noise on the source, and (ii) How can
the players adopt their actions if they have information on the
corruption rate of the source.

{\bf\textit{b. Eisert's scheme:}} In this physically realizable
scheme the quantum version of a two-player-two-strategy classical
game can be played as follows: (a) A referee prepares a maximally
entangled state by applying an entangling operator $\hat{J}$ on a
product state $|f\rangle|g\rangle$ where $\{f,g\}\in\{0,1\}$. The
output of this entangler,
\noindent  which reads
$\hat{J}|fg\rangle=\frac{1}{\sqrt{2}}[~|fg\rangle+i(-1)^{(f+g)}|(1-f)(1-g)\rangle~]$, is delivered to the players. (b) The players apply their actions,
which are SU(2) quantum operations locally on their qubits, and
return the resultant state
$|\phi\rangle_{\rm{out}}=(\hat{U}_{A}\otimes\hat{U}_{B})\hat{J}|fg\rangle$
back to the referee.  Operators $\hat{U}_{A}$ and $\hat{U}_{B}$
are restricted to two-parameter SU(2) operators given by
\begin{eqnarray}\label{N05}
\hat{U}=\left(%
\begin{array}{ccc}
  e^{i\phi}\cos\frac{\theta}{2} && \sin\frac{\theta}{2}\\
&\\
  -\sin\frac{\theta}{2} && e^{-i\phi}\cos\frac{\theta}{2} \\
\end{array}%
\right),
\end{eqnarray}
\noindent where $0\leq\phi\leq\pi/2$ and $0\leq\theta\leq\pi$. (c)
The referee, upon receiving this state, applies
$\hat{J}^{\dagger}$ and then makes a quantum measurement
$\Pi_{n}=|j \ell\rangle\langle j \ell|$ with $n=2j+\ell$ and
$j,\ell\in\{0,1\}$. Then the average payoffs of the players become
\begin{eqnarray}\label{N03}
&&\$_{A}=\sum_{n} a_{n}\underbrace{\rm{Tr}(\Pi_{n}\hat{J}^{\dagger}\hat{\rho}_{\rm{out}}\hat{J})}_{P_{j \ell}}\nonumber\\
&&\$_{B}=\sum_{n}
b_{n}\underbrace{\rm{Tr}(\Pi_{n}\hat{J}^{\dagger}\hat{\rho}_{\rm{out}}\hat{J})}_{P_{j
\ell}}
\end{eqnarray} where $\hat{\rho}_{\rm{out}}=|\phi\rangle_{\rm{out}}\langle \phi|$, $a_{n}$ and
$b_{n}$ are the payoffs chosen from the classical payoff matrix
when the measurement result is $n$, and $P_{j \ell}$ corresponds
to the probability of obtaining $n$. The classical version of the
game can be played using the same scheme if the operations
corresponding to the classical pure strategies are chosen as
$\hat{\sigma}_{0}$ and $i\hat{\sigma}_{y}$.

\begin{figure}[h]
\epsfxsize=8.7cm
\epsfbox{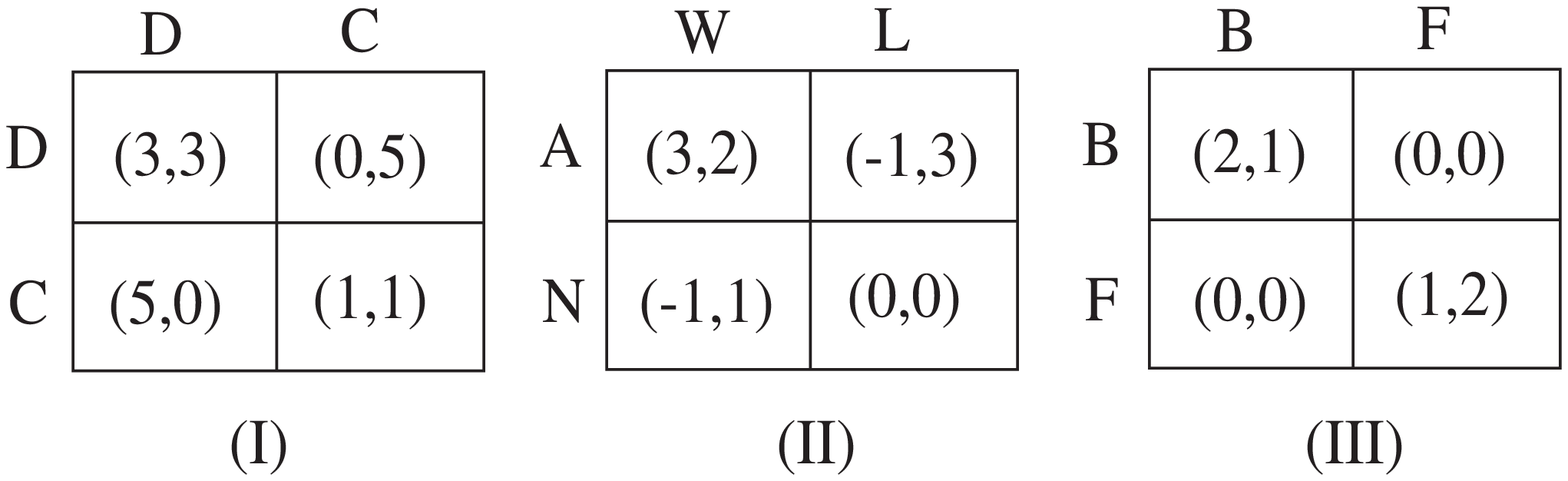}\vspace{-2mm}\caption[]{Payoff matrices for
the (I) Prisoner's dilemma (PD), (II) Samaritan's dilemma (SD),
and (III) Battle of Sexes (BoS). These games are $2\times2$ games,
that is each of the two-players- Alice (Column) and Bob (Row)- has
two possible actions. The action sets of the players are:  Deny
(D) and Confess (C) in PD; Work (W), Loaf (L), Aid (A), and No-aid
(N) in SD; Ballet (B), and Football (F) in BoS. The numbers in the
parenthesis denote the payoffs received by the players for their
action combinations. The first entry in the parenthesis is the
payoff for Alice, and the second one is that for
Bob.}\label{Fig:Payoffs}
\end{figure}

Using this scheme, quantum versions of some dilemma-containing
classical games, such as Prisoner's Dilemma (PD), Samaritan's
Dilemma (SD) and Battle of Sexes (BoS) whose payoffs matrices are
given in Fig.\ref{Fig:Payoffs}, have been studied. In these games,
it has been understood that if the referee starts with the state
$|fg\rangle=|00\rangle$ generating the entangled state
$|\Psi\rangle=\frac{1}{\sqrt{2}}[~|00\rangle+i|11\rangle~]$, the
players can resolve their dilemma and receive the highest possible
total payoff $\$_{A}+\$_{B}$. It has also been shown that the
dynamics of the games changes when the referee starts with a
different initial state. For example, if the referee starts with
$|fg\rangle=|01\rangle$ in SD, four NE's emerge with the same
constant payoff making a solution to the dilemma impossible
\cite{Ozdemir}.

\begin{figure}[h] \epsfxsize=7cm
\epsfbox{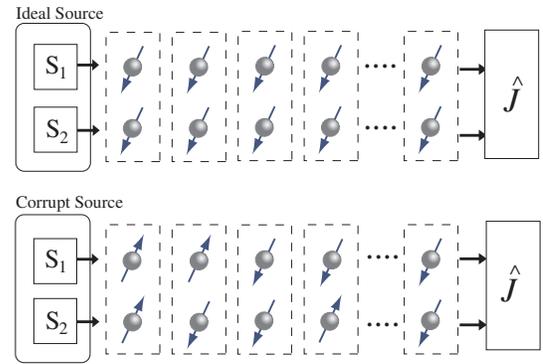}\caption[]{Model of the source for the
scheme in Fig. \ref{Fig:scheme}. The rounded rectangle, which
includes $S_{1}$ and $S_{2}$, is the source to prepare the initial
product state $|f\rangle|g\rangle$. In the ideal source, each of
$S_{1}$ and $S_{2}$ deterministically prepares the spin-down state
$|0\rangle$. Hence, at each run of the game, the prepared state is
$|f\rangle|g\rangle=|0\rangle|0\rangle$ with probability one. On
the other hand, in the corrupt source, $S_{1}$ and $S_{2}$ prepare
the spin-up state $|1\rangle$ with probability $r$, and the
spin-down state with probability $1-r$. Therefore, the input state
of the entangler, denoted by $\hat{J}$, becomes a mixture of
spin-up and spin-down states.}\label{Fig:corrupt}
\end{figure}
{\bf\textit{c. Corrupt Source in Quantum Games:}} As we have
pointed out above, the initial state from which the referee
prepares the entangled state is a crucial parameter in Eisert's
scheme. Therefore, any corruption or deviation from the ideality
of the source which prepares this state will change the dynamics
and outcomes of the game. Consequently, the analysis of situations
where the source is corrupt is necessary to shed a light in
understanding the game dynamics in the presence of imperfections.
We consider the source model shown in Fig. \ref{Fig:corrupt}. This
model includes two identical sources constructed to prepare the
states $|0\rangle$'s which are the inputs to the entangler at each
run of the game. These sources are not ideal and have a 
\emph{corruption rate}, $r$, that is, they prepare the
desired state $|0\rangle$ with probability $(1-r)$ while preparing
the unwanted state $|1\rangle$ with probability $r$. The state
prepared by these sources thus can be 
\begin{figure}[h]
\epsfxsize=6cm \epsfbox{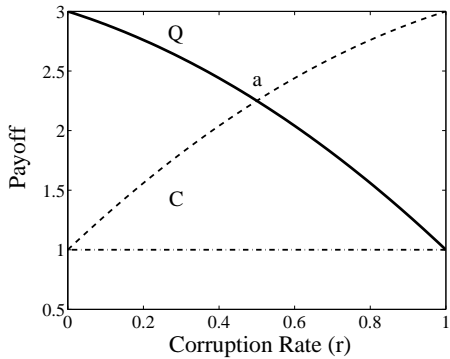}\caption[]{The payoffs received by
the players in Prisoner's Dilemma as a function of corruption
rate, $r$, for their quantum, $Q$, and classical, $C$, strategies.
The point labelled as $a$ at $r=1/2$ is the transition from
quantum advantage to classical advantage while the corruption rate
increases. The horizontal dashed line denotes the payoffs of both
players for the classical strategy without corruption.
}\label{Fig:PD}
\end{figure} 
\noindent written as
$\hat{\rho}_{1}=\hat{\rho}_{2}=(1-r)|0\rangle\langle
0|+r|1\rangle\langle 1|$. Then the combined state generated and
sent to the entangler becomes
$\hat{\rho}_{1}\otimes\hat{\rho}_{2}=(1-r)^{2}|00\rangle\langle
00|+r^{2}|11\rangle\langle 11|+r(1-r)(|01\rangle\langle
01|+|10\rangle\langle 10|)$. This  results in a mixture of the
four possible maximally entangled states
$(1-r)^{2}|\psi^{+}\rangle\langle\psi^{+}|+r^{2}|\psi^{-}\rangle\langle
\psi^{-}|+r(1-r)(|\phi^{+}\rangle\langle
\phi^{+}|+|\phi^{-}\rangle\langle \phi^{-}|)$, where
$|\psi^{\mp}\rangle=|00\rangle\mp i|11\rangle$ and
$|\phi^{\mp}\rangle=|01\rangle\mp i|10\rangle$. This is the state
on which the players will perform their unitary operators.

{\bf\textit{Scenario I:}} In this scenario, the players Alice and
Bob are not aware of the corruption in the source. They assume
that the source is ideal and always prepares the initial state $|fg\rangle=|00\rangle$, and hence that
the output state of the entangler is always $|\psi^{+}\rangle$.
Based on this assumption, they apply the operations that is supposed to resolve their dilemma.

We have analyzed PD, SD and BoS according to this scenario, and
compared the payoffs of the players with respect to the corruption
rate. The payoff they receive when they stick to their quantum
strategies are compared  to the payoffs when they play the game
classically. We consider  the classical counterparts both with and
without the presence of  noise in the game. That is, the players
use the  same physical scheme of the quantum version with and
without the corrupt source, and apply their actions by choosing
their operators from the set
$\{\hat{\sigma}_{0},i\hat{\sigma}_{y}\}$.

\begin{figure}[h]
\epsfxsize=6cm \epsfbox{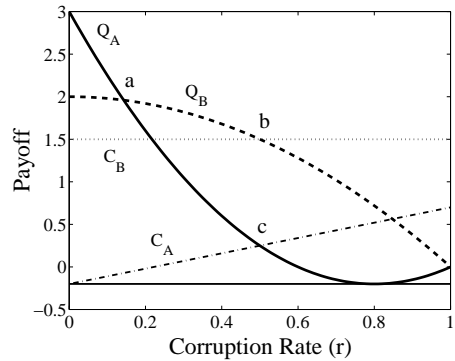}\vspace{-2mm}\caption[]{The
payoffs received by the players in Samaritan's dilemma as a
function of corruption rate, $r$, for their quantum, $Q_{A},
Q_{B}$, and classical, $C_{A},C_{B}$, strategies. The labelled
points are as follows: $a$, transition point from $\$_{A}>\$_{B}$
(\emph{Case 3} $\leftrightarrow$ \emph{Case 2}), $b$ and $c$ at
$r=1/2$, transitions from quantum advantage to classical advantage
for Bob and Alice, respectively, for increasing corruption rates.
The horizontal solid line denotes the payoff Alice receives in
classical strategies when the source is ideal. For Bob, classical
strategies with and without noise coincide and are depicted with
the horizontal dotted line.}\label{Fig:SD}
\end{figure}

\begin{figure}[h]
\epsfxsize=6cm \epsfbox{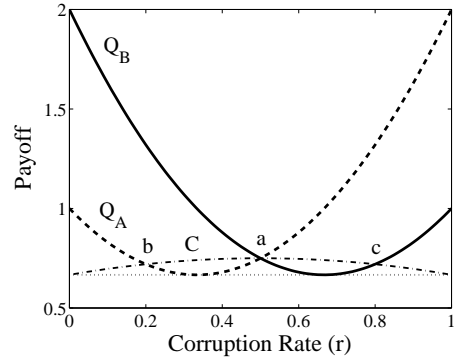}\vspace{-2mm}\caption[]{The
payoffs received by the players in Battle of Sexes as a function
of corruption rate, $r$, for their quantum, $Q_{A}, Q_{B}$, and
classical, $C$, strategies. For the increasing corruption rate,
the transition from quantum to classical advantage occurs at
points labelled as $b$ for $r=0.2$ and $a$ for $r=1/2$,
respectively, for Alice and Bob. The second transition that is a
transition from the classical to quantum advantage occurs at $a$
for $r=1/2$ and $c$ for $r=0.8$, respectively, for Alice and Bob.
The point labelled as $a$ is also the transition point from
$\$_{B}>\$_{A}$ to $\$_{A}>\$_{B}$. The horizontal dotted line
corresponds to the payoffs of the players for their classical
strategy when the source is ideal. }\label{Fig:BoS}
\end{figure}

The results of the analysis according to this scenario  are
depicted in Figs. \ref{Fig:PD}-\ref{Fig:BoS}. A remarkable  result
of this analysis is that with the introduction of the corrupt
source, the players' quantum advantage is no longer preserved  if
the corruption rate, $r$, becomes larger than a critical
corruption rate $r^{\star}$. At $r^{\star}$, the classical and
quantum strategies produce  equal payoffs. Another interesting
result is the existence of a strategy
$\hat{U}_{A}=\hat{U}_{B}=(\hat{\sigma}_{0}+i\hat{\sigma}_{y})/\sqrt{2}$,
where the payoffs of the players become constant independent of
corruption rate. 
\noindent This strategy could be attractive
for risk avoiding and/or paranoid players.

For PD, which is a symmetric game, the optimal classical
strategies deliver the payoffs $\$_{A}=\$_{B}=1$ for the actions
$\hat{U}_{A}=\hat{U}_{B}=i\hat{\sigma}_{y}$. In the quantum
version with an uncorrupt source,  the players can get the optimal
payoffs $\$_{A}=\$_{B}=3$ if they adopt the strategies
$\hat{U}_{A}=\hat{U}_{B}=i\hat{\sigma}_{z}$  \cite{Eisert1}. 
Hence, the dilemma of
the game is resolved and the players receive better payoffs than
those obtained with classical strategies.  However, as seen in
Fig. \ref{Fig:PD}, the payoffs of players with classical and
quantum strategies become equal to $2.25$ when $r=r^{\star}=1/2$. If $r$
satisfies $0\leq r<1/2$, the quantum version of the game always
does better than the classical one. Otherwise, the classical game
is better.

When the classical version of PD is played with a corrupt source,
we find that with increasing corruption rate, while the payoffs
for the quantum strategy decrease,  those of the classical one
increase. That is, if $r>r^{\star}$,  then the players would
rather apply their classical strategies  than the quantum ones.
This can be explained as follows: When the  players apply
classical operations, the game is played as if  there is no
entanglement in the scheme. That is, players apply  their
classical operators $i\hat{\sigma}_{y}$ on the state prepared by  the
source. If the source is ideal, $r=0$, they operate on the
$|00\rangle$ which results in an output state $|11\rangle$.
Referee, upon receiving this output state and making the
projective measurement, delivers $\$_{A}=\$_{B}=1$. On the other
hand, when $r=1$, the  state from the source is $|11\rangle$ and
the output state after  the players actions becomes $|00\rangle$.
With this output state,  referee delivers them the payoffs
$\$_{A}=\$_{B}=3$. Thus, when the players apply the classical
operator  $i\hat{\sigma}_{y}$, their payoffs continuously increase
from one to three with the increasing corruption rate from $r=0$
to $r=1$.

 Using a classical mixed strategy in the asymmetric game of SD,
 the players receive $(\$_{A},\$_{B})=(-0.2,1.5)$ at the NE. In this
 strategy, while Alice chooses from her strategies with equal
 probabilities, Bob uses a biased randomization where he applies one of his actions, $\hat{\sigma}_{0}$, with probability
 $0.2$. The most desired solution to the dilemma in the game is to
 obtain an NE with $(\$_{A},\$_{B})=(3,2)$. This is achieved when both players
 apply $i\hat{\sigma}_{z}$ to $|\psi^{+}\rangle$ \cite{Ozdemir}. The dynamics of the payoffs in this game with the corrupt source when
 the players stick to their operators $i\hat{\sigma}_{z}$ and its
 comparison with their classical mixed strategy are depicted in
 Fig. \ref{Fig:SD}. Since this game is an asymmetric one, the
 payoffs of the players, in general, are not equal. However, with the corrupt source it is found that
 their payoffs become equal at $r=1/7$ and at $r=1$, where the payoffs are $96/49$ and $0$, respectively.
 The critical corruption rate, $r^{\star}=1/2$, which denotes the
transition from the quantum advantage to classical advantage
regions, is the same for both players. While for increasing $r$,
$\$_{B}$ monotonously decreases from two to zero, $\$_{A}$ reaches
its minimum of $-0.2$ at $r=0.8$, where it starts increasing to
the value of zero at $r=1$. It is worth noting that when the
players apply their classical mixed strategies in this physical
scheme, $\$_{B}$ is always constant and independent of the
corruption rate, whereas $\$_{A}$ increases linearly  as
$\$_{A}=-0.2+0.9r$ for $0\leq r\leq 1$. The payoffs of the players
are compared in three cases \cite{Ozdemir}: \emph{Case 1}:
$\$_{A}\leq0$ (insufficient solution), \emph{Case 2}:
$0<\$_{A}\leq\$_{B}$ (weak solution), and \emph{Case 3}:
$0\leq\$_{B}<\$_{A}$ (strong
 solution). In the corrupt source scenario in quantum strategies, \emph{Case 1} is achieved for
 $0.6\leq r\leq 1$,\emph{Case 2} is achieved for $1/7\leq r< 0.6$,
 and finally \emph{Case 3} for $r< 1/7$. The remarkable result of
 this analysis is that although the players using quantum strategies have high potential gains,
 there is a large potential loss if the source is deviated from an
 ideal one. The classical strategies are more robust to corruption
 of the source.

 In BoS, which is an asymmetric game, the classical mixed strategies, where Alice and Bob apply $\hat{\sigma}_{0}$ with probabilities $1/3$ and $2/3$
 or vice versa, the players
 receive equal payoffs of $2/3$. However, the dilemma is not solved due to the
existence of two equivalent NE. On
 the other hand, when the physical scheme with quantum strategies
 is used the players can reach an NE where their payoffs
 become $\$_{A}=1$ and $\$_{B}=2$ if both players apply
$i\hat{\sigma}_{y}$ to the maximally entangled state prepared with
an ideal source \cite{Note1}. The advantage of this quantum
strategy to the classical mixed strategy is that in the former
$\$_{A}+\$_{B}$ is higher than the latter. In the presence of
corruption in the source, payoffs of the players change as shown
in Fig. \ref{Fig:BoS}. With an ideal source, the payoffs reads
$(\$_{A},\$_{B})=(1,2)$, however for increasing corruption rate
while $\$_{B}$ decreases from two to one, 
\noindent $\$_{A}$ increases from
one to two. With an completely corrupt source, $r=1$,
 the payoffs become $(\$_{A},\$_{B})=(2,1)$. The reason for this is the same as
explained for PD. When the quantum strategies with and without
corrupt source are compared to the classical mixed strategy
without noise, it is seen that the former ones always give better
payoffs to the players. However, when the source becomes noisy
(corrupt), classical strategies become more advantageous to quantum ones with
increasing corruption rate. The range of corruption rate where
classical strategies are better than the quantum strategy if the
players stick to their operations $i\hat{\sigma}_{y}$, are
$0.2<r<0.5$ and $0.5<r<0.8$ for Alice and Bob, respectively. When
$r=1/2$, $\$_{A}=\$_{B}$ and these payoffs are equal to the ones
received with classical mixed strategies. While $\$_{A}=\$_{B}$
independently of $r$ for classical mixed strategies, $\$_{A}$ and
$\$_{B}$ differ when the players stick to their quantum strategies
for $r\neq1/2$. Another interesting result for this game is that,
contrary to PD and SD, the strategy
$\hat{U}_{A}=\hat{U}_{B}=(\hat{\sigma}_{0}+i\hat{\sigma}_{y})/\sqrt{2}$
discussed above always gives a constant payoff $(3/4,3/4)$, which
is better than that of the classical mixed strategy.

{\bf\textit{Scenario II:}} In this scenario, the referee knows the
characteristics of the corruption in the source, and inform the
players on the corruption rate. Then the question is whether the
players can find a unique NE for a known source with corruption
rate $r$; and if they can, does this NE resolve their dilemma in
the game or not. When the corruption rate is $r=1/2$, the state
shared between the players become $\hat{\rho}=\hat{I}/4$. Then
independent of what action they choose, the players receive
constant payoffs determined by averaging the payoff entries 
\begin{table}[h]
\begin{center}
\begin{tabular}{ccccccc}
\hline &$r$&$~~~~\hat{U}_{A}(\theta_{A},\phi_{A})~~~~$&$~~~~\hat{U}_{B}(\theta_{B},\phi_{B})~~~~$&$(\$_{A},\$_{B})$ \\
\hline
&$0$& $~~~~(0,\frac{\pi}{2})$&$(0,\frac{\pi}{2})$&$(3,3)$   \\
&$1/4$&$~~~~(0,\frac{\pi}{2})$&$(0,\frac{\pi}{2})$&$(\frac{43}{16},~\frac{43}{16})$   \\
&$1/2$& $~~~~(\theta_{A},\phi_{A})$\footnotemark[1]&$(\theta_{B},\phi_{B})$\footnotemark[1]&$(\frac{9}{4},~\frac{9}{4})$   \\
&$3/4$&$~~~~(0,\frac{\pi}{4})$&$(0,\frac{\pi}{4})$&$(\frac{43}{16},~\frac{43}{16})$   \\
&$1$&  $~~~~(0,\frac{\pi}{4})$&$(0,\frac{\pi}{4})$&$(3,3)$  \\
\hline
\end{tabular}
\footnotetext[1]{~$\forall \theta_{A},\forall \theta_{B} \in
[0,\pi]$ and $\forall \phi_{A},\forall \phi_{B} \in [0,\pi/2]$ }
\caption{Strategies which lead to NE's and the corresponding payoffs
for the players in Prisoner's Dilemma (PD) if they are provided
the information on the corruption rate, $r$, of the source which
prepares in the initial product state $|0\rangle|0\rangle$.
\label{tab:prob1}}
\end{center}
\end{table}
\begin{table}[h]
\begin{center}
\begin{tabular}{ccccccc}
\hline &$r$&$~~~~\hat{U}_{A}(\theta_{A},\phi_{A})~~~~$&$~~~~\hat{U}_{B}(\theta_{B},\phi_{B})~~~~$&$(\$_{A},\$_{B})$ \\
\hline
&$0$&  $~~~~(0,\frac{\pi}{2})$&$(0,\frac{\pi}{2})$&$(3,2)$   \\
&$1/4$& $~~~~(0,\frac{\pi}{2})$&$(0,\frac{\pi}{2})$&$(\frac{21}{16},~\frac{15}{8})$   \\
&$1/2$& $~~~~(\theta_{A},\phi_{A})$\footnotemark[1]\footnotemark[2]&$(\theta_{B},\phi_{B})$\footnotemark[1]\footnotemark[2]&$(\frac{1}{4},~\frac{3}{2})$   \\
&$3/4$& $~~~~(0,\phi)$\footnotemark[2]&$(0,\frac{\pi}{2}-\phi)$\footnotemark[2]&$(\frac{21}{16},~\frac{15}{8})$   \\
&$1$& $~~~~(0,\phi)$\footnotemark[3]&$(0,\frac{\pi}{2}-\phi)$\footnotemark[3]&$(3,2)$   \\
\hline
\end{tabular}
\footnotetext[1]{~$\forall \theta_{A},\forall \theta_{B} \in
[0,\pi]$} \footnotetext[2]{~$\forall \phi_{A},\forall \phi_{B} \in
[0,\pi/2]$} \footnotetext[3]{~$\phi\in [0,\pi/4]$}
\caption{Strategies which lead to NE's and the corresponding payoffs
for the players in Samaritan's Dilemma (SD) if they are provided
the information on the corruption rate, $r$, of the source which
prepares in the initial product state $|0\rangle|0\rangle$.
\label{tab:prob2}}
\end{center}
\end{table}
\noindent
in the classical game payoff matrices. In this case, the players get equal payoffs $9/4$ and $3/4$ for PD and BoS, respectively. In SD,
the payoffs are $\$_{A}=1/4$ and $\$_{B}=3/2$.

For PD, an interesting result is that there is no difference in
the payoffs between an ideal source, $r=0$, and a completely
corrupt source, $r=1$. That is, the players can resolve the
dilemma receiving the best possible payoffs, $\$_{A}=\$_{B}=3$, in
both cases. However, the strategies which lead to a unique NE in
these two extreme cases are different: When $r=0$, the
players can resolve the dilemma by applying
$\hat{U}_{A}=\hat{U}_{B}=i\hat{\sigma}_{z}$; however when $r=1$,
they have to change their actions to
$\hat{U}_{A}=\hat{U}_{B}=(\hat{\sigma}_{0}+i\hat{\sigma}_{z})/\sqrt{2}$
in order to resolve the dilemma.

For BoS, while an NE is achieved resolving the dilemma with
$\$_{A}+\$_{B}=3$ for both $r=0$ and $r=1$, the corruption rate
shows its effect in the payoffs and the actions to reach NE's.
When $r=0$, the payoffs are $(\$_{A},\$_{B})=(1,2)$, on the other
hand when $r=1$, payoffs become $(\$_{A},\$_{B})=(2,1)$. As can be
seen in Table \ref{tab:prob3}, the difference in the strategies is
the choice of $\phi_{A}$ and $\phi_{B}$; while for $r=1$ the
players should choose $\phi_{A}=\phi_{B}=0$ to arrive at the NE,
for $r=0$ they have an infinite number of choices for $\phi_{A}$
and $\phi_{B}$ and any of these choices will work equally well.

The effect of a corrupt source is much stronger 
\begin{table}[h]
\begin{center}
\begin{tabular}{ccccccc}
\hline &$r$&$~~~~\hat{U}_{A}(\theta_{A},\phi_{A})~~~~$&$~~~~\hat{U}_{B}(\theta_{B},\phi_{B})~~~~$&$(\$_{A},\$_{B})$ \\
\hline
&$0$& $~~~~(\theta,\phi)$\footnotemark[1]&$(\theta,\frac{\pi}{2}-\phi)$\footnotemark[1]&$(1,2)$   \\
&$~$& $~~~~(\pi,\phi_{A})$\footnotemark[2]&$(\pi,\phi_{B})$\footnotemark[2]&$(1,2)$   \\
&$1/4$& $~~~~(\theta,\phi)$\footnotemark[1]&$(\theta,\frac{\pi}{2}-\phi)$\footnotemark[1]&$(\$_{A}^{'},\$_{B}^{'})$  \\
&$~$& $~~~~(\pi,\phi_{A})$\footnotemark[2]&$(\pi,\phi_{B})$\footnotemark[2]&$(\frac{11}{16},\frac{19}{16})$   \\
&$1/2$& $~~~~(\theta_{A},\phi_{A})$\footnotemark[2]\footnotemark[3]&$(\theta_{B},\phi_{B})$\footnotemark[2]\footnotemark[3]&$(\frac{3}{4},~\frac{3}{4})$   \\
&$3/4$&$~~~~(\pi,0)$&$(\pi,0)$&$(\frac{19}{16},\frac{11}{16})$    \\
&$1$& $~~~~(\pi,0)$&$(\pi,0)$&$(2,1)$    \\
\hline \end{tabular} \footnotetext[1]{~$\theta\in [\pi/2,\pi]$ and
$\phi\in [0,\pi/2]$}\footnotetext[2]{~$\forall \phi_{A},\forall
\phi_{B} \in [0,\pi/2]$}\footnotetext[3]{~$\forall
\theta_{A},\forall \theta_{B} \in [0,\pi]$}\caption{Strategies
which lead to NE's and the corresponding payoffs for the players
in BoS if they are provided the information on the corruption rate
of the source which prepares in the initial product state
$|0\rangle|0\rangle$. Note that one of the sets of strategies
leading to NE's when $r=1/4$ gives payoffs depending on $\theta$.
These payoffs are as follows: $\$_{A}'=(13-2\cos(2\theta))/16$ and
$\$_{B}'=(20-\cos(2\theta))/16$.  \label{tab:prob3}}
\end{center}
\end{table}
\noindent for the SD game.
In this game, in contrast to the other two, although for $r=0$
there is a unique NE solving the dilemma, for $r=1$ the players
cannot find a unique NE. There emerges an infinite number of
different strategies with equal payoffs $(3,2)$. The players are
indifferent between these strategies and cannot make sharp
decisions. Therefore, the dilemma of the game survives, although
its nature changes.

When we look at some intermediate values for the corruption rate,
we see that corruption rate affects BoS and SD strongly. For
example, when $r=3/4$ in SD, there are infinite number of strategies and NE's which have the
same payoffs $(21/16,15/8)$. These NE's are achieved when the
players choose their operators as $\theta_{A}=\theta_{B}=0$ and
$\phi_{B}=-\phi_{A}+\pi/2$. The same is seen in BoS for $r=1/4$
which results in a payoff $(15/16,21/16)$ when the players choose
$\theta_{A}=\theta_{B}=\pi/2$ and $\phi_{B}=-\phi_{A}+\pi/2$. A
more detailed analysis carried out for PD with increasing $r$ in
steps of $0.1$ in the range $[0,1]$ has revealed that the players
can achieve a unique NE where their payoffs and strategies depends
on the corruption rate. Therefore, information on the source
characteristic might help the players to reorganize their
strategies. However, whether providing the players with this kind
of information in a game  is acceptable or not is an open
question.

{\bf\textit{d. Conclusion:}} This study shows that the strategies
to achieve NE's and the corresponding payoffs are strongly
dependent on the corruption of the source. In a game with corrupt
source, the quantum advantage no longer survives if the corruption
rate is above a critical value. The corruption may not only cause
the emergence of multiple NE's but may cause a decrease in the
player's payoff, as well, even if there is a single NE. If the
players are given the characteristics of the source then they can
adapt their strategies; otherwise they can either continue their
best strategy assuming that the source is ideal and take the risk
of losing their quantum advantage over the classical or choose a
risk-free strategy, which makes their payoff independent of the
corruption rate. However, in the case where players know the
corruption rate and adjust their strategies, the problem is that
for some games there emerge multiple NE's, therefore the dilemmas
in those games survive. This study reveals the importance of the
source used in a quantum game.

\begin{acknowledgments}
The authors thank to Dr. J. Soderholm for the critical reading of
the manuscript. They also acknowledge the insightful discussions
with Dr. F. Morikoshi and Dr. T. Yamamoto.
\end{acknowledgments}

\end{document}